\documentclass[runningheads]{llncs}
\usepackage{graphicx}
\usepackage{svg}
\usepackage{xcolor}
\begin{document}
\title{Through their eyes: multi-subject Brain Decoding with simple alignment techniques}
\titlerunning{Cross-subject Brain Decoding}
%
\author{Matteo Ferrante\inst{1} \and
Tommaso Boccato\inst{1}\and
Nicola Toschi \inst{1,3}}
\authorrunning{Ferrante et al.}
%
\institute{Department of Biomedicine and Prevention \\ University of Rome Tor Vergata (IT)
\email{matteo.ferrante@uniroma2.it}\\
 \and 
 Martinos Center For Biomedical Imaging \\ MGH and Harvard Medical School (USA)
}
\maketitle              
\begin{abstract}
To-date, brain decoding literature has focused on single-subject studies, i.e. reconstructing stimuli presented to a subject under fMRI acquisition from the fMRI activity of the same subject. The objective of this study is to introduce a generalization technique that enables the decoding of a subject's brain based on fMRI activity of another subject, i.e. cross-subject brain decoding. To this end, we also explore cross-subject data alignment techniques. Data alignment is the attempt to register different subjects in a common anatomical or functional space for further and more general analysis.

We utilized the Natural Scenes Dataset, a comprehensive 7T fMRI experiment focused on vision of natural images. The dataset contains fMRI data from multiple subjects exposed to 9841  images, where 982 images have been viewed by all subjects. Our method involved training a decoding model on one subject's data, aligning new data from other subjects to this space, and testing the decoding on the second subject based on information aligned to first subject. We also compared different techniques for fMRI data alignment, specifically ridge regression, hyper alignment, and anatomical alignment.

We found that cross-subject brain decoding is possible, even with a small subset of the dataset, specifically, using the common data, which are around $10\%$ of the total data, namely 982 images, with performances in decoding compararble to the ones achieved by single subject decoding. Cross-subject decoding is still feasible using half or a quarter of this number of images with slightly lower performances. Ridge regression emerged as the best method for functional alignment in fine-grained information decoding, outperforming all other techniques.

By aligning multiple subjects, we achieved high-quality brain decoding and a potential reduction in scan time by $90\%$. This substantial decrease in scan time could open up unprecedented opportunities for more efficient experiment execution and further advancements in the field, which commonly requires prohibitive (20 hours) scan time per subject.

\end{abstract}
\section{Introduction}

\begin{figure}[t]
    \centering
    \includegraphics[width=\textwidth]{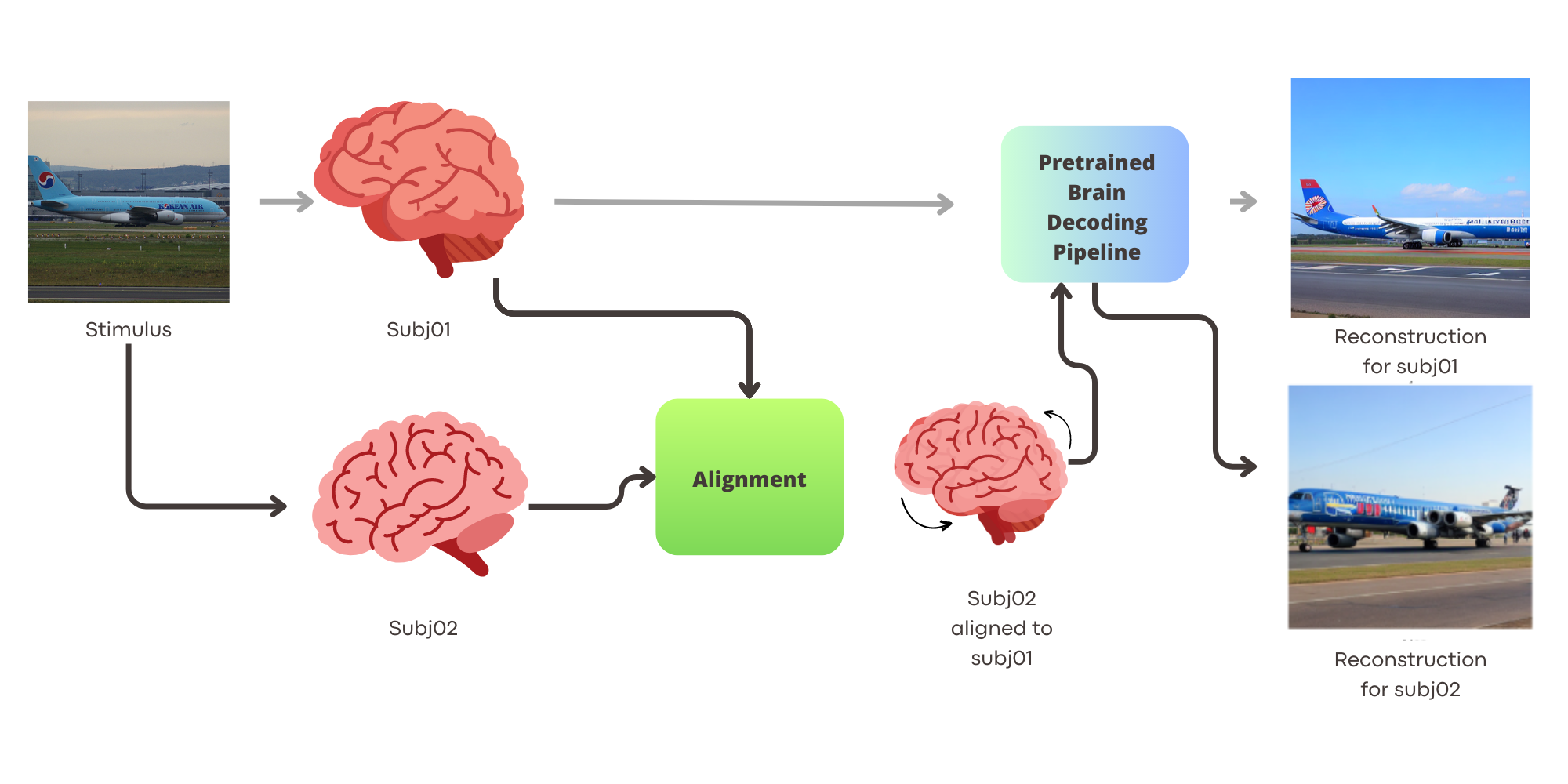}
    \caption{Scheme for cross-subject decoding: The procedure involves the following steps: In the first step, Subj01 (top row) is selected as the target subject. A decoding model is trained to reconstruct seen images based on the brain activity of Subj01 on the training set of Subj01 images (8859 images). These images showed uniquely to this subject.
    Next, we decode the brain activity of a second subject, Subj02, who was exposed to a share of the same stimuli that Subj01 was exposed to (982 images).  Using the shared images we can compare the brain activity related to the same stimuli across different subjects. We used this shared information to align the functional activity of Subj02 with that of the target subject. Once the alignment transformation is learned, we can align the complete dataset (including unseen data) and we can utilize the pretrained decoder to reconstruct images from Subj02 without training a decoding model specifically for Subj02.}
    \label{fig:cross_pipeline}
\end{figure}

Deep learning has revolutionized numerous fields, including neuroscience. The application of deep learning techniques in neuroscience has led to significant advancements in understanding brain function and decoding the intricate workings of the human mind. Brain decoding, in particular, has emerged as a crucial area where deep learning plays a pivotal role.

Brain decoding involves the extraction of meaningful information from recorded brain activity, allowing researchers to infer mental states, perceptual experiences, or cognitive processes. For example, deep learning algorithms have been used to decode brain activity and predict whether an individual is looking at a face or an object based on their neural responses \cite{zafar_decoding_2015,awangga_literature_2020}.

The potential applications of brain decoding are vast, from understanding various aspects of brain function, such as information processing strategies, decision-making, memory formation, and consolidation, to potential uses in neurofeedback, neuroaesthetics, or neuromarketing strategies \cite{neurostuff}. Moreover, successful brain decoding could lead to novel strategies for diagnosing and treating neurological or neuropsychiatric conditions, and potentially contribute to the development of radically new algorithmic learning strategies.
However, these promising endeavors are not without challenges. Noninvasive data, for instance, have lower temporal or spatial resolution than neural firing, which may limit the granularity of information that can be retrieved. Furthermore, physiological noise and signal/image artifacts can affect both fMRI and EEG data, which can only be imperfectly removed after the data are acquired. Nevertheless, several brain decoding studies have achieved impressive results \cite{zafar_decoding_2015,awangga_literature_2020}. 

One of the key challenges in brain decoding is the subject-specific nature of all models developed thus far. This means that the models are tailored to individual subjects, which can lead to significant variability in the results, given that the amount of data collection per subject could be limited by external factors like time and acquisition costs. Moreover, intrinsic inter-individual variability poses further challenges and every model has to be built from scratch for each new subject. This variability is a consequence of the unique functional and anatomical structure of each individual's brain, and implies the need to  acquire an entire dataset and train an individual model for each subject. This technology's use is limited by a bottleneck requiring extensive data collection—typically thousands of stimulus images—to function properly. This complexity stems from the unique brain anatomy, information processing methods, and functional responses each individual possesses, complicating the training of a universally applicable brain activity decoding model.
Despite individual differences in brain anatomy and function, common structures enable reliable neuroscience analysis using template matching techniques, such as anatomical and functional alignment. Anatomical alignment transforms individual brain images to match a standard 'average' brain template or 'atlas', aligning the size, shape, and orientation of brain images. This facilitates meaningful cross-comparisons of brain images, although it is more effective for larger, well-defined structures and may lack precision for smaller, variable regions. Additionally, it does not account for functional differences across brains. Thus, anatomical alignment is often supplemented with functional alignment, which synchronizes brain activity patterns across individuals, aiding the comparison and analysis of functional data. This method is vital as activity locations can differ among individuals. Numerous functional alignment methods exist, each with distinct applications and limitations.
\begin{figure}[!ht]
    \centering
    \includegraphics[width=\textwidth]{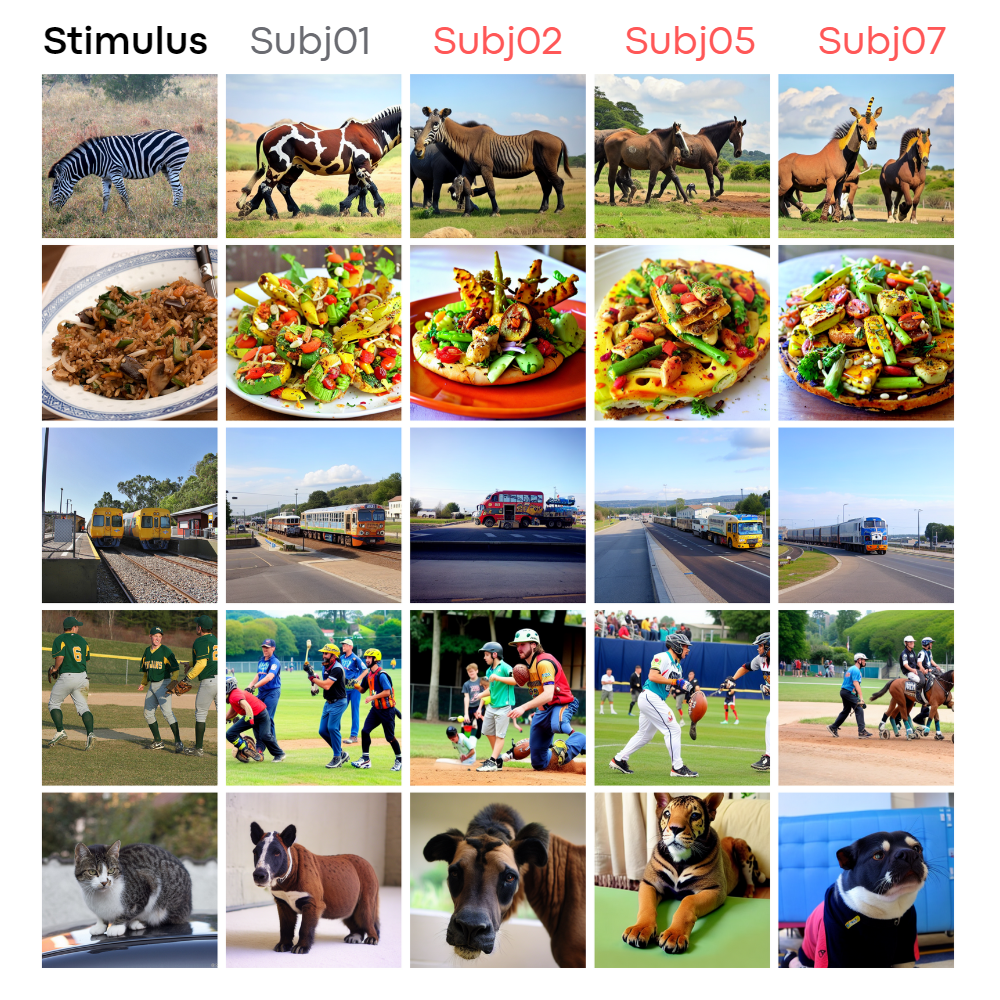}
    \caption{Example results: The first column, "Stimulus", presents the stimuli from the fMRI experiment. The "Subj01" column (in gray) displays the decoded activity from Subj01, providing an upper performance baseline using a subject-specific decoder model. All other columns (in red) show results from functional alignment using Ridge Regression with $100\%$ common data (952 images), meaning subjects were functionally aligned to Subj01 and decoded using Subj01's trained decoder. To ensure robust visual comparisons, none of the displayed images were used in learning alignment transformation, demonstrating functional alignment on unseen data not used for decoder training or alignment function learning.}
    \label{fig:ridge_big_1}
\end{figure}

In this study, we explore and contrast three methods for cross-subject brain decoding of visual stimuli, using the established, cutting-edge Brain-Diffuser decoding procedure. This approach mitigates variability from new decoding procedures, enabling straightforward quantitative and qualitative comparisons of cross-subject decoding outcomes.

We train a visual stimuli decoding model for one subject (Subj01) and employ anatomical alignment, hyper alignment, and functional alignment with ridge regression for three others, decoding their activity using the pretrained model. Our goal is to demonstrate the feasibility of fine-grained cross-subject decoding for visual stimuli reconstruction, potentially reducing scan times significantly by only acquiring data necessary for alignment, thereby reaching state-of-the-art in image reconstruction. Figure \ref{fig:cross_pipeline} outlines our proposed pipeline, while Figure \ref{fig:ridge_big_1} provides examples of cross-subject decoding results.

\section{Related Work}
\begin{figure}
    \centering
    \includegraphics[width=\textwidth]{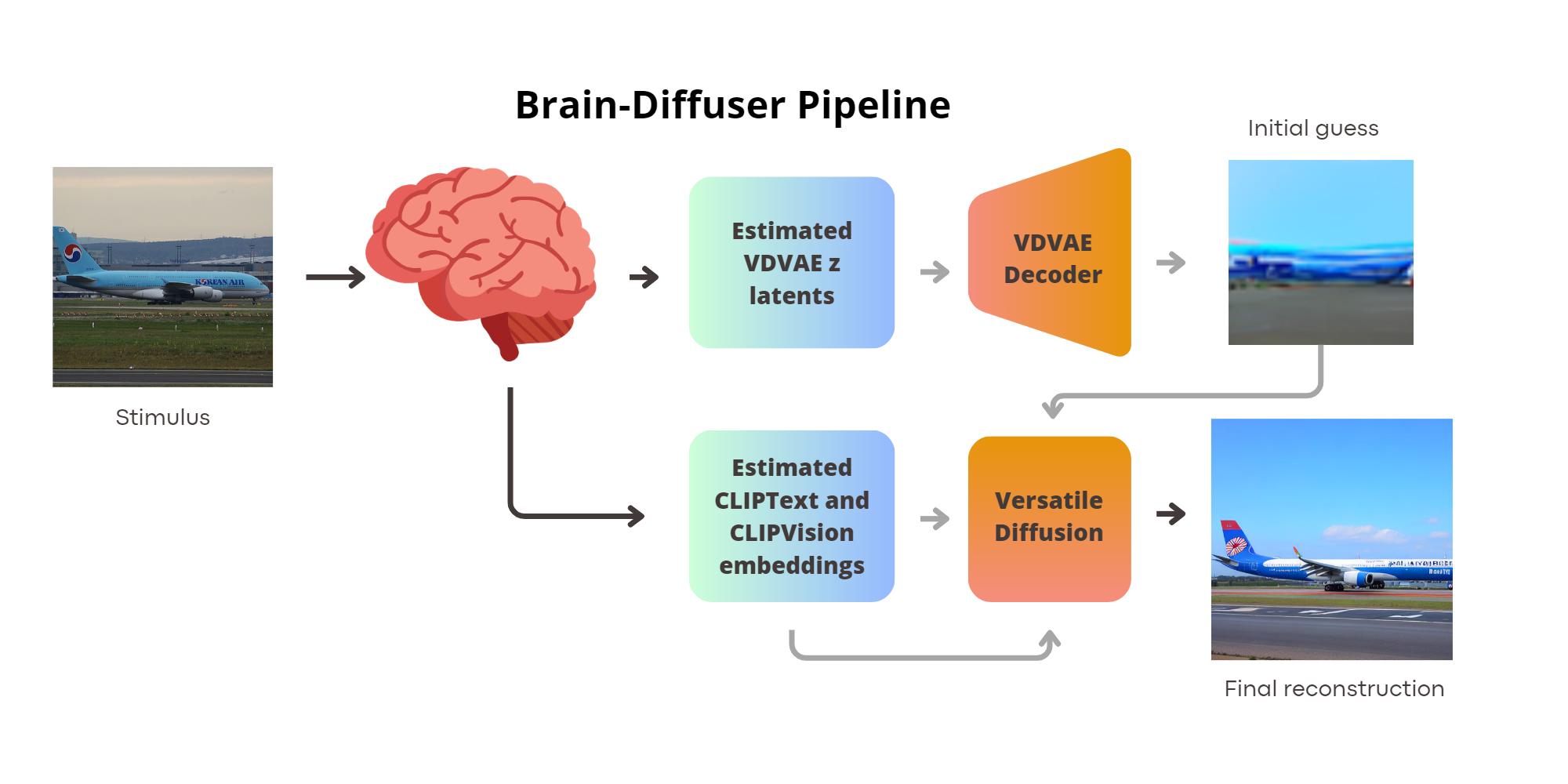}
    \caption{The Brain-Diffuser pipeline, the state-of-the-art decoder for brain activity used in this study, begins with brain activity from viewing an image stimulus. A model is trained to estimate the latent representation of the VDVAE autoencoder as well as the text and visual embeddings of the CLIP model, using linear models. These estimated vectors and an initial guess image obtained by decoding the autoencoder latents, are fed into Versatile Diffusion—a latent diffusion model—to reconstruct the final image.}
    \label{fig:braindiff_pipeline}
\end{figure}

In the burgeoning field of deep learning-based brain decoding, a range of models has been utilized to scrutinize preprocessed fMRI time series as input, particularly focusing on visual stimuli decoding. This involves reconstructing images that could have triggered specific fMRI patterns—termed brain activity. Here, we review major works in this domain.

Some methods have employed variational autoencoders with a generative adversarial component (VAE-GAN) to encode latent human face representations, estimating these encoded representations from fMRI activity using a linear model \cite{faces}. Sparse linear regression has also been utilized on preprocessed fMRI data to predict features from the early convolutional layers of a pre-trained convolutional neural network (CNN) for natural images \cite{horikawa_generic_2017}.

Unsupervised and adversarial strategies have been used for image reconstruction, including dual VAEGAN and unsupervised methods for decoding fMRI stimuli, utilizing multiple encoder and decoder approaches \cite{shen_end--end_nodate,ren_reconstructing_2019,gaziv_self-supervised_2022}. Pretrained architectures like BigBiGAN and IC-GAN have optimized latent spaces, significantly enhancing high-fidelity image reconstruction from fMRI patterns \cite{bigbigan,icgan}.

Recently, diffusion models have become prominent in the decoding pipeline, providing superior image generation performance \cite{Takagi2022.11.18.517004,chen2022seeing}. These models often incorporate semantic-based strategies \cite{ferrante2023semantic} and multi-step decoding strategies \cite{ozcelik2023braindiffuser,ferrante2023brain}.

Regarding alignment techniques, there are several approaches \cite{functional_review}. Hyperalignment \cite{Haxby2011-up,hyper} aligns functional brain activity across individuals in a high-dimensional space, enhancing the precision of cross-subject brain activity predictions, but requires extensive high-quality data and complex computational resources. The Shared Response Model (SRM) \cite{SRM} aligns brain activity by identifying a common response pattern across subjects, ideal for shared experiences, but assumes uniform responses, which individual perception and cognition differences may contradict. Independent Component Analysis (ICA) \cite{ica}, separates multivariate signals into additive subcomponents, identifying common brain activity patterns, but requires statistical independence of subcomponents, which may not always apply to brain data.

While functional alignment methods provide powerful tools for comparing brain activity, their limitations and assumptions require careful result interpretation. These methods align and compare functional brain data, complementing, not replacing, anatomical alignment. Various other methods, each with its pros and cons, have been proposed.

In this paper, we compare anatomical alignment, hyperalignment-based functional alignment, and ridge regression-based alignment methods for cross-subject brain decoding.

\section{Material and Methods}

\begin{figure}[t]
    \includegraphics[width=\textwidth]{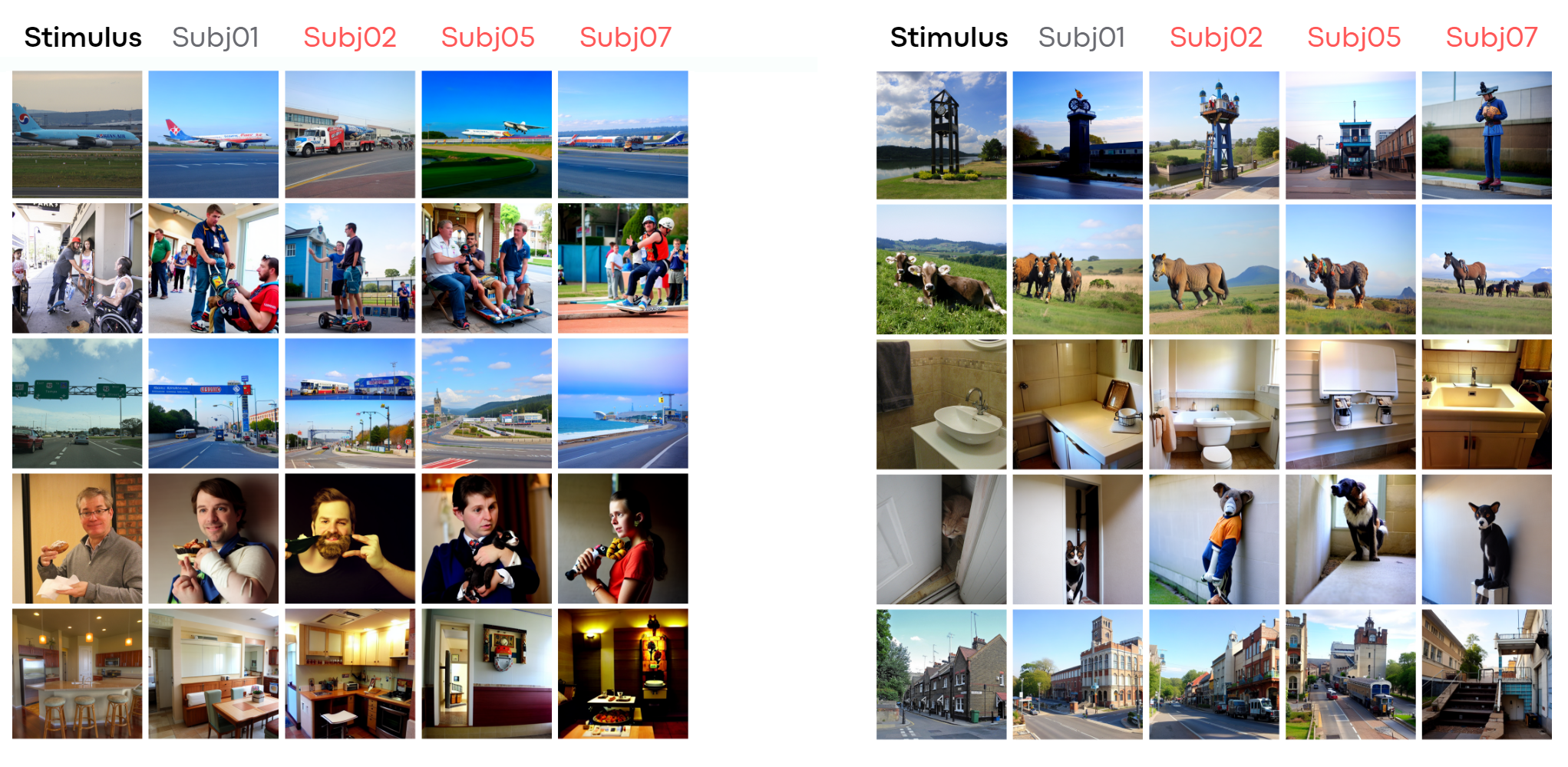}
    \caption{More example results. Format and conventions as in  Figure 2}
    \label{fig:comparison_1}
\end{figure}
In this section, we describe the proposed method and the data we used. The data are publicly available and can be requested at \url{https://naturalscenesdataset.org/}. All experiments and models were trained on a server equipped with four NVIDIA A100 GPU cards (80GB RAM each connected through NVLINK) and 2 TB of System RAM.

The study utilizes the Natural Scenes Dataset (NSD) \cite{NSDDataset}, a vast fMRI data set from eight subjects exposed to images from the COCO21 dataset. We focused on four subjects, forming a unique training dataset of 8,859 images and 24,980 fMRI trials, and a common dataset of 982 images and 2,770 trials. To reduce spatial dimensionality, we applied a mask to the fMRI signal (resolution of 1.8mm isotropic) using the NSDGeneral ROI, targeting various visual areas. This strategic ROI selection enhanced the signal-to-noise ratio and simplified data complexity, enabling exploration of both low-level and high-level visual features. Temporal dimensionality was reduced using precomputed betas from a general linear model (GLM) with a fitted hemodynamic response function (HRF) and a denoising process as detailed in the NSD paper. Data from Subj01, Subj02, Subj05, Subj07, warped into the Montreal Neurological Institute common space (MNI) and downsampled at 2mm, represented the brain activity of each subject and helped decrease computational time and cost.
We used the common dataset as alignment, keeping out 30 images for visual comparison, so there are 8859 unique images for each subject. We only used them for training the decoding model for Subj01. Then there are 952 common images across all subjects that was used to functionally align them to the activity of Subj01, and 30 common images kept out for visual comparison on images neither used in the training or in the alignment procedure. These 30 images were chosen because they're used as visual qualitative evaluation of decoding results in other papers and could help the reader to compare results across different methods. Decoding metrics are evaluated on the 952 images which correspond to our test set for each one of the subjects, since these images are never seen by the decoder model, so the evaluation is still fair and on unseen images. When we refer to $100\%$ of common data we are pointing to these 952 images.

\subsection{Decoding Pipeline: Brain-Diffuser}

The "Brain-Diffuser" \cite{ozcelik2023braindiffuser} model is a two-stage framework for reconstructing natural scenes from fMRI signals. Initially, a Very Deep Variational Autoencoder (VDVAE) provides an "initial guess" of the reconstruction, focusing on low-level details. This guess is refined using high-level semantic features from CLIP-Text and CLIP-Vision models, employing a latent diffusion model (Versatile Diffusion) for final image generation. The model, represented in Fig. \ref{fig:braindiff_pipeline}, takes fMRI signals as input and generates reconstructed images, capturing low-level properties and overall layout. As a state-of-the-art procedure, Brain-Diffuser was trained using data from Subj01 in the MNI space (cross-subject decoding requires a of a common space). Further details about the decoding model are available in the original paper.

\begin{figure}[t]
\centering
\includegraphics[width=\textwidth]{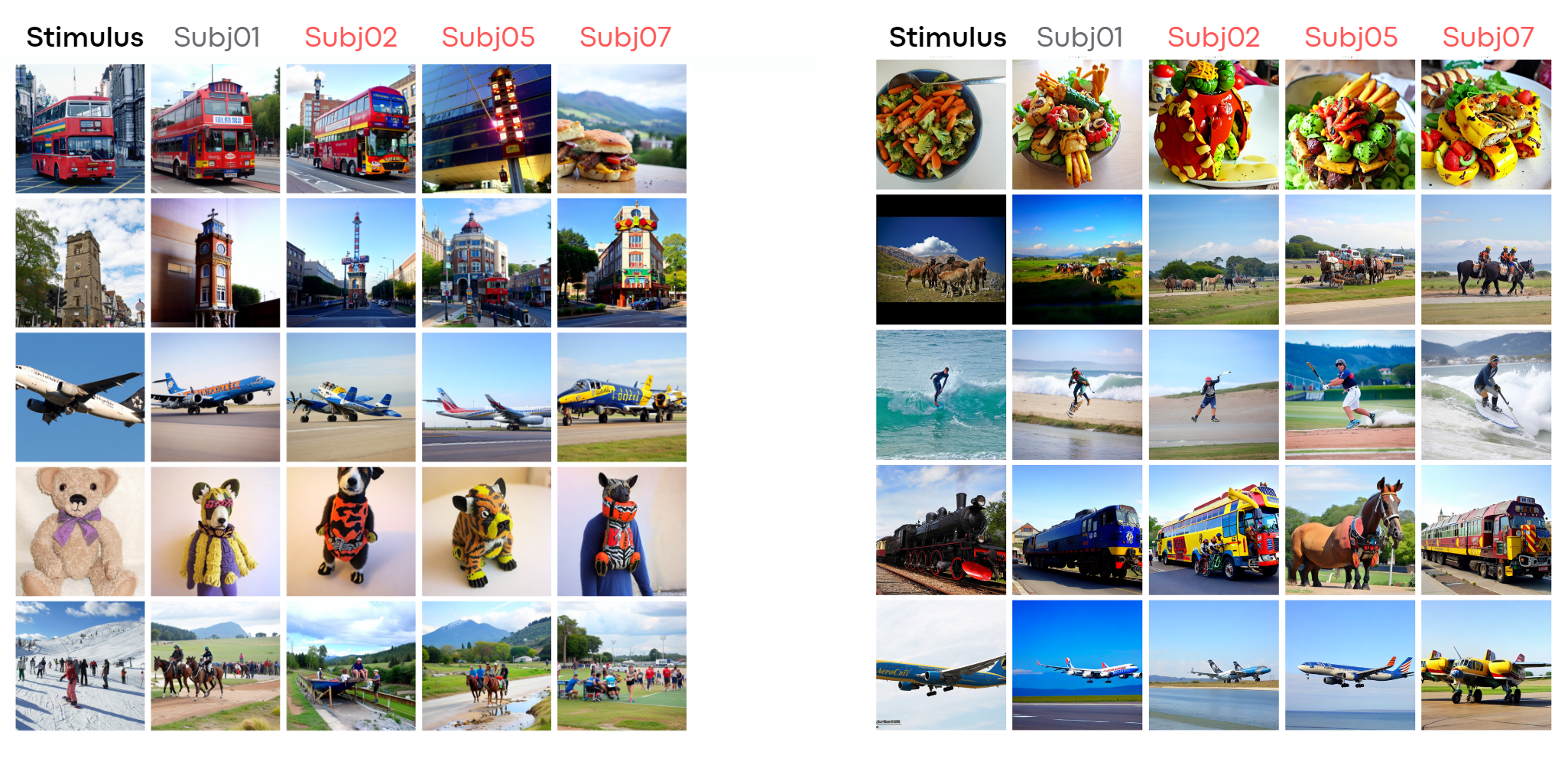}
\caption{More example results. Format and conventions as in  Figure 2}
\label{fig:comparison_2}
\end{figure}

\subsection{Alignment}
This study investigates three alignment strategies to evaluate cross-subject fine-grained brain decoding's feasibility: anatomical alignment, functional alignment via hyper alignment, and functional alignment through ridge regression.

Anatomical alignment, our baseline, relies solely on anatomical details, transforming functional aspects using pre-computed structural image transformations. On the other hand, functional alignment necessitates a more comprehensive approach. Consider the scenario where the brain activity of a source subject $\mathbf{S}$ needs to align with a target subject $\mathbf{T}$. These activities, responses to numerous stimuli, are matrices of shape \textit{($\#$ stimuli, $\#$ voxels)}. Given that subjects encounter several common stimuli (i.e., they view identical images in the fMRI scanner), we can divide the datasets into $\mathbf{T}{common}, \mathbf{T}{different}$ and $\mathbf{S}{common}, \mathbf{S}{different}$. Our goal is to leverage the $common$ dataset portion to learn a mapping from $S$ to $T$, aligning the entire $\mathbf{S}$ dataset with the $\mathbf{T}$ functional space. The NSD experiment's structure, with separate training and test sets (the latter containing identical images for each subject), provides a common stimuli set for alignment purposes.

\subsubsection{Anatomical Alignment}
Anatomical alignment, a common neuroscience method, aligns to a standard template, here, the MNI space, facilitating anatomical structure comparison. This alignment typically involves a linear coregistration of anatomical images between native and common spaces, followed by a nonlinear warping to match common brain structures. Several software options like FSL and ANTs can perform this task. The NSDData authors \cite{NSDDataset} elaborate on this process in their released code, providing betas (i.e. coefficients obtained by theressing the stimuls waveform against the fMRI data) for all subjects in the MNI common space (1mm). We downsampled these to 2mm to approximate the resolution used in the original Brain-Diffuser decoding paper (1.8mm) and to reduce spatial dimensionality.

\subsubsection{HyperAlignment}
HyperAlignment \cite{Haxby2011-up,hyper}, a functional data alignment technique, models functional data as high-dimensional points, with each voxel representing a dimension with betas ranging in $\mathcal{R}$. This method, based on Procrustes Analysis \cite{Gower1975}, presents a high-dimensional model of the representational space in the human ventral temporal (VT) cortex, wherein dimensions are response-tuning functions common across individuals.

To perform the Procrustes analysis for functional brain alignment, we aim to find a rotation matrix $\mathbf{R}$ and a scale factor $c$ such that the difference $|c\mathbf{S}\mathbf{R} - \mathbf{T}|^2$ is minimized.

This is achieved by computing the matrix product $\mathbf{P}=\mathbf{S}^T_{common} \mathbf{T}_{common}$,
Performing the singular value decomposition of $\mathbf{P}$ to obtain left and right eigenvector matrices $\mathbf{U}$ and $\mathbf{V}$,
Computing $\mathbf{R}=\mathbf{U}\mathbf{V}^T$ and the scaling factor $c=\frac{trace(\mathbf{T_{common}}^T (\mathbf{S_{common}} \mathbf{R}))}{trace(\mathbf{S}^T_{common}\mathbf{S}_{common})}$.
Finally, we can apply the matrix $\mathbf{R}$ and the scaling $c$ to both common and non-common source data to align them with the target subject. We computed these values for Subj02, Subj05, and Subj07 as source subjects, using Subj01 as the target, to align all subjects to the functional space of the first one. For detailed mathematical proofs and other insights, please refer to the original articles \cite{hyper,Haxby2011-up,Gower1975}.

\begin{figure}[!h]
    \centering
    \includegraphics[width=\textwidth]{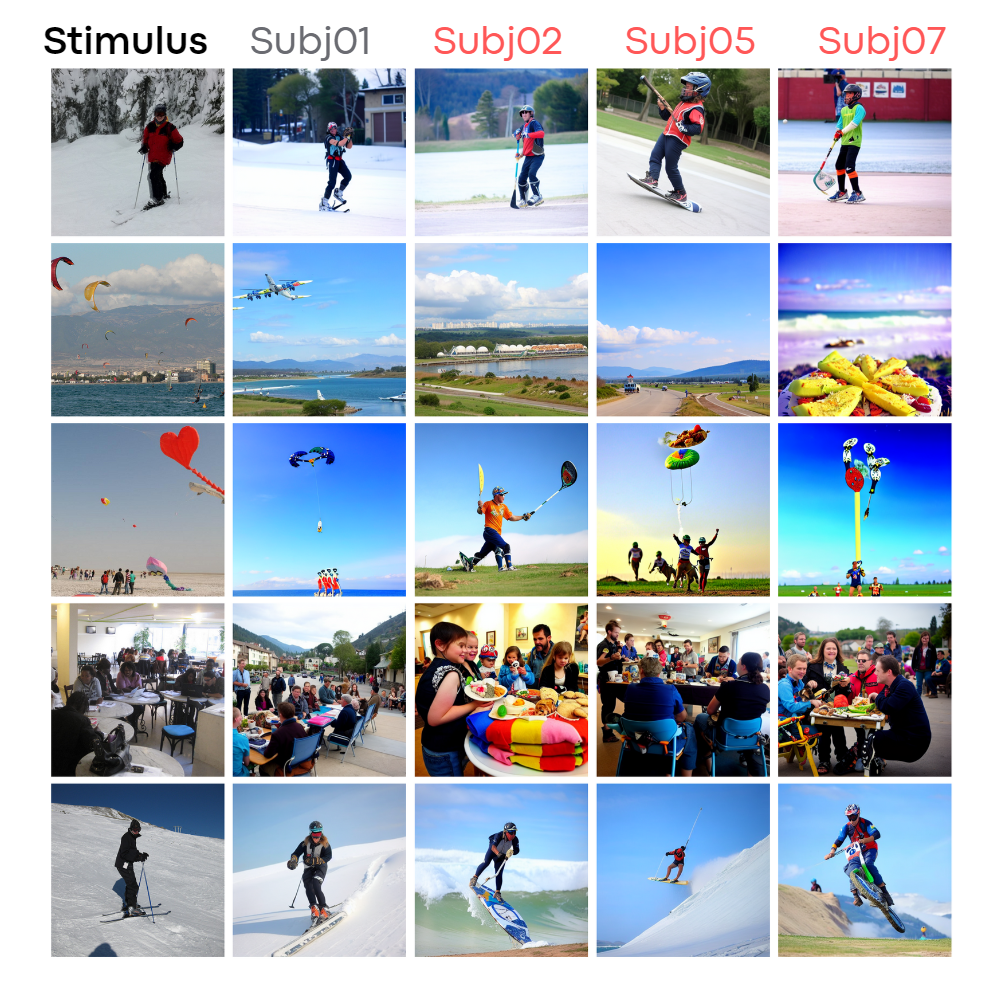}
    \caption{More example results. Format and conventions as in  Figure 2}
    \label{fig:ridge_big_2}
\end{figure}

\subsubsection{Ridge Regression}
Our third approach embraces a simple assumption: even in different subjects, all functional data contain the information for the same stimuli, albeit possibly spread across different voxels. This suggests that one subject's activity (source) might be expressed as a linear combination of the activity of another subject (target) for the same stimuli. By deriving a linear combination for each voxel of the target from all possible voxels of the source, we can create a linear map from the source to the target, facilitating functional alignment. The target subject activity can be expressed as $\mathbf{t}i=\sum{j} \mathbf{w}{j}\mathbf{s}{j}$, where $\mathbf{t}i$ is the $i$-th activity of the target voxel for each common dataset stimulus. Here, $\mathbf{t}i$ represents the $i$-th column of $\mathbf{T}{common}$, expressed as a linear combination of all $\mathbf{S}{common}$ columns. The challenge lies in finding the vector of $\mathbf{w}$ values. When extended to all the target subject voxels, the $w$ vector morphs into a square matrix $\mathbf{W}$, each column of which contains weights to estimate one target subject voxel from a combination of source values. The objective can be redefined as minimizing $|\mathbf{S}{common}\mathbf{W}^T-\mathbf{T}{common}|^2$.

We employed Ridge Regression from \cite{sklearn_api} to determine the $\textbf{W}$ matrix, conducting a 5-fold cross-validation to select the optimal hyper-parameter $\alpha$ from the values $[0, 1, 10, 1e2,1e3,1e4]$. Our findings indicated that $\alpha=1000$ yields superior performance, hence we adopted it as our final regularization parameter. We computed these values to align all subjects to the initial functional space For the sources Subj02, Subj05, and Subj07, and using Subj01 as the target.

\subsection{Evaluation}
Our research seeks to evaluate visual stimuli's detailed brain decoding feasibility, scrutinizing the alignment methods and shared data ratio at play. We examined how the alignment performance fluctuates when the shared data makes up $10\%$, $25\%$, $50\%$, and $100\%$ of the total common data (952 images).

Our shared dataset, or "test dataset," comprises 982 images, all viewed by every subject. In order to allow visual comparison, we excluded 30 images from the original Brain-Diffuser paper. Thus, these excluded images neither influenced the training of the decoding pipeline nor the alignment process. The remaining 952 images serve as the shared dataset. We computed transformations for each alignment method (anatomical, hyperalignment, ridge regression) and shared dataset proportion, applying the linear transformation to the complete dataset. This procedure aligns the images with Subject 01's functional space. We then used the pre-trained Brain-Diffuser pipeline for decoding the aligned fMRI activity and reconstructing the images. We assessed our image reconstruction process through both basic and advanced metrics, including PixCorr, SSIM, and 2-way accuracy in AlexNet, Inception, and CLIP latent spaces. This comprehensive evaluation approach allows us to benchmark our results against other brain decoding studies. However, the goal here is not merely comparison, but rather the examination of performance in relation to the shared data fraction and alignment method, given a fixed decoding pipeline, trained solely on Subj01 as a reference target.

\section{Results}

\begin{figure}[t]
\centering
\includegraphics[width=\textwidth]{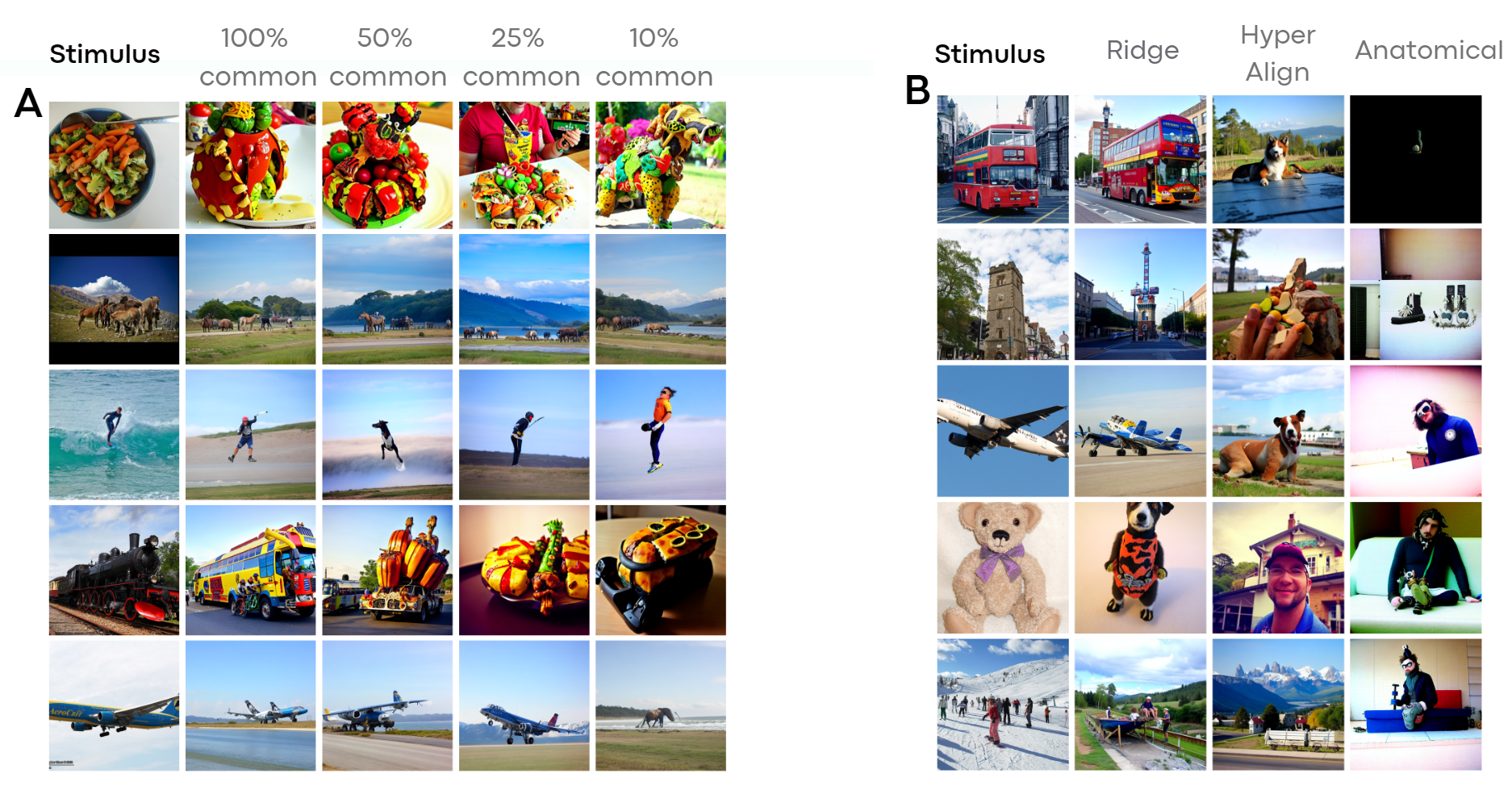}
\caption{\textbf{A}: This figure compares Ridge Regression functional alignment over different fractions of common data. The "Stimulus" column presents the images used during the experiment, with the remaining columns illustrating the decoded, aligned activity of Subj02 on Subj01. \textbf{B}: Here, we compare various alignment modalities. Again, the "Stimulus" column displays the images used during the experiment, while the other columns visualize the decoded activity of Subj02 aligned to Subj01 using different methods.}
\label{fig:comparison_3}
\end{figure}

\begin{figure}[!h]
\centering
\includegraphics[width=\textwidth]{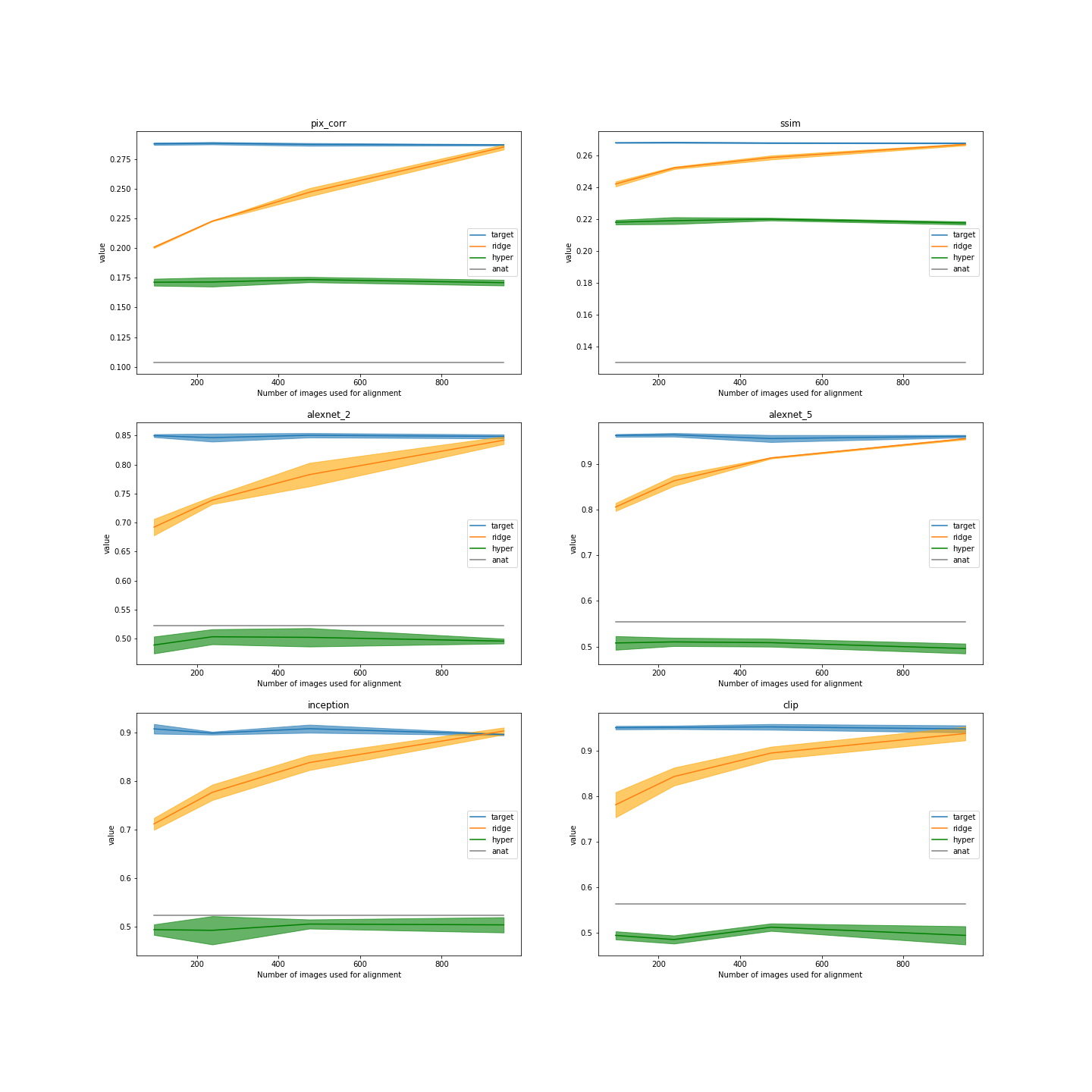}
\caption{This figure illustrates the performance of various methods evaluated using different metrics. Blue lines represent metrics from the target subject's decoded images, derived from their test set brain activity. Green lines denote the mean and standard deviation (std) of performance on test sets from other subjects, aligned using hyperalignment. Gray lines present results achieved using anatomical alignment, while orange lines display outcomes using Ridge Regression. Remarkably, the Ridge Regression approach yields positive results even when using a tiny fraction of the entire dataset. Furthermore, as this fraction approaches roughly $10\%$ of the total set, resulting in 952 images the performance becomes comparable with those obtained by the within-subject model.}
\label{fig:metrics}
\end{figure}

Figures \ref{fig:comparison_1}, \ref{fig:comparison_2}, \ref{fig:ridge_big_1}, \ref{fig:ridge_big_2} provide a comparison between stimuli and decoded images from Subj01 (on which the decoding model is trained). These figures also display the aligned activity of all other subjects using Ridge Regression. Figure \ref{fig:comparison_3} compares fractions of common data used for Ridge Regression-based alignment and different alignment methods. Lastly, Figure \ref{fig:metrics} illustrates each quantitative metric, computed and averaged over the entire test set for each aligned subject (2,5,7). Metrics are expressed as a fraction of the entire dataset, which contains approximately 10k images per subject (8859 unique + 982 common across subjects). As common images are necessary, the maximum amount of images that can be included in the alignment process is 952 (termed the "test set", except 30 images left out for visualization purposes), representing around 10\% of the dataset. This represents the maximum data that can be incorporated into the procedure, and we experimented with half, a quarter, and a tenth of this data.

Anatomical and Hyperalignment methods fail to yield satisfactory results, demonstrating just above chance performance levels for 2-way classification accuracy and poor performance for low-level metrics such as SSIM and PixCorr. However, Ridge Regression exhibits an increasing performance based on the volume of data used for alignment mapping function learning. This method reaches performance levels comparable with the within-subject decoder in both low-level and high-level metrics, using all the common data (approximately 10\% of the entire dataset).

Our findings are encapsulated in the following key points:

\textbf{Functional alignment's critical role in fine-grained brain decoding}: Our research emphasizes the pivotal role of functional alignment in fine-grained brain decoding. This process, which interprets neural signals to reconstruct perceived images or thoughts, greatly benefits from precise functional alignment of brain activity. Accurate alignment ensures that neural signals are matched correctly to their corresponding brain regions, thus enhancing the decoding accuracy.

\textbf{Anatomical method's inefficacy}: As corroborated by previous studies \cite{hyper}, our research found that anatomical methods for brain decoding are ineffective. Relying on the physical structure of the brain for alignment and decoding does not deliver the requisite precision for fine-grained decoding tasks. This could be attributed to inherent anatomical variability across different individuals, which may not necessarily align with functional differences. The specialized areas in the brain with functional selectivity can sometimes yield performance above chance levels. However, in most cases, decoded images do not correlate with the stimulus, undermining the reliability of this method for cross-subject brain decoding.

\textbf{Overfitting tendency of complex techniques}: We noted that more sophisticated techniques, like hyperscanning for brain decoding, tend to overfit the data. This results in poor generalization to unseen data, with metrics measuring n-way accuracy reaching only chance levels. While these techniques might seem to offer superior decoding accuracy initially, their lack of generalizability limits their practical utility. Of course, room for improvement exists, perhaps through the incorporation of regularization techniques.

\textbf{Ridge Regression's efficacy}: Our results demonstrate that Ridge Regression-based methods for brain decoding can achieve above-chance performance levels with as little as 1\% of the entire dataset. Furthermore, these methods near baseline performance levels with around 10\% of the dataset. This crucial finding implies that reliable brain decoding results can be achieved while significantly reducing scan time. This efficiency could be instrumental in making brain decoding research and applications more feasible and cost-effective.

These results contribute to our comprehension of the challenges and potential remedies in brain decoding and emphasize the need for additional research to refine these techniques and augment their effectiveness.

\section{Discussion}

Our study underscores the intricacies and potential of cross-subject fine-grained brain decoding, a field promising to enhance our understanding of the human brain and cognition.

We identified the criticality of functional alignment for successfully executing brain decoding. This alignment, which maps neural signals to their corresponding brain regions, is vital for accurately decoding neural activity from other individuals using a model pre-trained on separate subject data. This insight holds promise for constructing large studies with a high-accuracy decoding pipeline, subsequently requiring only alignment data acquisition for new subjects. This approach negates the need for an entire experimental reproduction each time, streamlining the process.

Our research reveals the limitations of anatomical methods for brain decoding, which rely on the physical brain structure for alignment and decoding. These methods underperformed due to inherent brain anatomical variability across individuals, which may not align with functional differences. Thus, this study emphasizes the need for functional, not merely anatomical, considerations in decoding studies.

Excitingly, our results suggest significant reductions in scan-time are possible. Ridge Regression-based methods were found to provide reliable brain decoding results with just a fraction of the entire dataset, implying practical implications for brain decoding research feasibility and cost-effectiveness.

Our study also highlights the qualitative similarities in decoded images across subjects. While these images largely match high-level semantic content, intra-subject differences appear minimized. This observation prompts us to consider whether the decoding procedure is fully captured. Given that high-level concepts are generally aligned, we propose a possible brain activity decomposition into \textit{brain activity} $=$ \textit{concept} + \textit{individual perception}. Such a model might only capture the \textit{concept} while treating differences as noise, offering new research directions to explore fine-grained inter-subject differences.

Apart from speculation, our findings suggest that despite individual brain structure and function differences, common neural activity patterns can be decoded across individuals. This opens up intriguing opportunities for developing generalized brain decoding models applicable across different individuals. However, our study also reveals the limitations of current functional alignment methods. Future research might consider exploring more complex models, such as neural networks, known for their ability to capture intricate, non-linear relationships, potentially improving functional alignment.

Looking ahead, several promising future work directions emerge, such as training models across subjects and machines, which could lead to more robust and generalizable brain decoding models. The development of new techniques and methodologies could potentially address current limitations of brain decoding, heralding more accurate and efficient decoding of brain activity. As we delve into fine-grained brain decoding, addressing potential privacy concerns and ethical implications is paramount. Current research suggests that while certain brain activity aspects can be decoded across subjects, the process is not yet a comprehensive or intrusive 'mind-reading' tool. A key finding highlights the disruptive role of attention mechanisms, suggesting that brain decoding is only possible with actively participating, aware subjects.

While our methods currently prevent involuntary or covert 'mind reading', as the field advances, maintaining strong ethical frameworks for brain decoding research becomes even more critical. Informed consent, strict data privacy protocols, and potential societal implications consideration remain key. Decoding brain activity raises broader ethical questions, such as its potential use to enhance communication for individuals with speech or motor impairments or its potential misuse for coercive or manipulative purposes. These critical questions must be confronted by the scientific community and society as we continue to explore brain decoding potential.

\section{Conclusions}

In this study, we present a method for conducting brain decoding of visual stimuli across different subjects. We delved into the role of functional alignment in decoding neural signals with accuracy, uncovering the limitations inherent in anatomical methods and the pitfalls of complex decoding techniques that are susceptible to overfitting.

Our study saw a significant breakthrough with the successful implementation of Ridge Regression-based methods. This technique showed remarkable efficiency in decoding neural activity, utilizing only a fraction of the dataset. This discovery indicates the possibility of substantial reductions in scan time - with potential reductions nearing 90$\%$. Such an advancement could revolutionize the feasibility and cost-effectiveness of brain decoding research and its applications.

In our research, we achieved successful cross-subject brain decoding by training the decoding pipeline on one subject and effectively decoding neural activity across multiple individuals. This significant finding points to the existence of shared neural activity patterns, paving the way for the development of generalized brain decoding models. We also revealed a hierarchical structure in the brain's processing and representation of information, separating brain decoding into concept and perception components. Despite these advancements, we recognize the limitations of current functional alignment methods and advocate for exploring future research directions, such as training models across subjects and machines.

In conclusion, our study illuminates the challenges and potential solutions within the sphere of fine-grained cross-subject brain decoding, particularly in relation to visual stimuli.
%
%
%
\bibliographystyle{splncs04}
\bibliography{ref}

\end{document}